# Computational Doppler-limited dual-comb spectroscopy with a free-running all-fiber laser

Łukasz A. Sterczewski[1], Aleksandra Przewłoka[2], Wawrzyniec Kaszub[2], Jarosław Sotor[1,*]

[1]Laser & Fiber Electronics Group, Faculty of Electronics, Wroclaw University of Science
and Technology, Wybrzeze Wyspianskiego 27, 50-370 Wroclaw, Poland
[2]Institute of Electronic Materials Technology, Wolczynska 133, 01-919 Warsaw, Poland
Author e-mail address: jaroslaw.sotor@pwr.edu.pl

**Abstract:**
Dual-comb spectroscopy has emerged as an indispensable analytical technique in applications that require high resolution and broadband coverage within short acquisition times. Its experimental realization, however, remains hampered by intricate experimental setups with large power consumption. Here, we demonstrate an ultra-simple free-running dual-comb spectrometer realized in a single all-fiber cavity suitable for the most demanding Doppler-limited measurements. Our dual-comb laser utilizes just a few basic fiber components, allows to tailor the repetition rate difference, and requires only 350 mW of electrical power for sustained operation over a dozen of hours. As a demonstration, we measure low-pressure hydrogen cyanide within 1.7 THz bandwidth, and obtain better than 1% precision over a terahertz in 200 ms enabled by a drastically simplified all-computational phase correction algorithm. The combination of the unprecedented setup simplicity, comb tooth resolution and high spectroscopic precision paves the way for proliferation of frequency comb spectroscopy even outside the laboratory.

## Introduction

Molecular spectroscopy with optical frequency combs (OFC) has experienced a remarkable growth in the last decade propelled by the brilliant concept of the simultaneously broadband and discrete nature of the light source, allowing to merge high-resolution laser spectroscopy with broadband spectral coverage. The relentless pursuit for precision[1] drove the scientific community to progressively turn initially free-running mode-locked lasers into precise frequency rulers[2–6] referenced to atomic clocks with relevance for astrophysics studies[7] in search for exoplanets.

New avenues for OFC-based spectroscopy have been opened with the inception of the dual-comb (DCS) technique[8–10] employing optical beating between two repetition rate mismatched frequency combs on a photodetector, which quickly surpassed the speed and precision of conventional Fourier Transform Spectroscopy (FTS). The novel all-solid-state spectroscopic technique paved the way for a highly-miniaturized real-time optical spectrum analyzer that multi-heterodynes discrete comb teeth to establish a direct link between the optical and radio frequencies (RF) domains, permitting acquisition times ranging from micro to milliseconds. In the last years, a surge of activity in the development of chip-scale dual-comb systems provided compact DCS platforms based on microresonators[11–13], quantum cascade[14,15], and interband cascade lasers[16], which hold a promise for a portable, field-deployable spectroscopic sensor covering different spectral regions.

Unfortunately, the convenience of DCS has its price. A majority of dual-comb spectrometers relies on two OFCs synchronized using complex phase locking loops necessitating the use of additional single-mode lasers, photodetectors, and fast control electronics[4]. Otherwise, substantial amounts of phase noise arising from unsynchronized operation of two free-running oscillators smear out the discrete character of the RF lines, hence precluding reliable spectroscopic assessments. To circumvent this issue, a well-known noise suppression technique can be employed, conceptually similar to balanced transmission widely used in electrical engineering. Because in DCS a critical requirement is mutual rather than absolute stability, by generating two combs in a single cavity, most of the environmental perturbations causing phase-noisy operation of the combs becomes common, hence such combs show a much higher degree of mutual coherence than a pair generated in two independent cavities. While their absolute frequency remains unstabilized, from a practical standpoint optical frequency drifts occurring on timescales suitable for spectroscopy with (sub)percentage level of uncertainty are much lower than molecular Doppler linewidths, which makes them compatible even with demanding high-resolution applications.

The burgeoning field of single-cavity dual-comb (SCDC) lasers is constantly being enriched with new concepts. To date, the most popular fiber-based SCDC generation mechanisms employ bidirectional mode-locking of a ring cavity[17–20], dual-wavelength lasing[21–25], and non-linear pulse shaping mechanisms[26]. Their full spectroscopic potential, however, has been weakly exploited to date. A vast majority reported on measurements at atmospheric pressure with apodized spectra, where gigahertz resolution is more than sufficient, while their repetition rates were relatively low – they ranged between a dozen and several dozen of megahertz. To date, such DCS systems have been perceived as suitable only for coarse spectroscopic investigations[4]. A notable exception from the above was provided by Zhao et al., who demonstrated picometer resolution measurements of acetylene using an all-fiber dual-color mode-locked laser[21]. Although impressive, the experiment employed nonlinear broadening mechanisms relying on a chain of optical components and pumping diodes for optical amplification of spectrally-filtered sub-combs, which adds an extra layer of complexity and may fail to meet tight power budget constraints. Similar issues impede a rivaling electro-optic (EOM) comb generation technique, which despite the repetition rate agility and ease of optical frequency axis calibration[27] also requires nonlinear broadening following intricate microwave electrooptic circuits for competitive spectral coverage. Unexpectedly, even such custom-synthesized and intrinsically mutually coherent sources may require phase correction techniques for DCS[28].

In our work, we show for the first time that by employing polarization multiplexing[29–32] in a drastically simplified all-fiber single-cavity dual-comb laser composed of just a few basic fiber components, one can perform high-resolution broadband molecular spectroscopy with tooth-resolved rf lines using a completely free-running system characterized by low power consumption and compact size. To our best knowledge, this is the simplest configuration for all-fiber dual-comb generation presented to date because two collinearly overlapping pulse trains with different repetition rates and high degree of mutual coherence are directly accessible from the tap of the output coupler without any additional optical or electronic components. Our demonstration opens up a new horizon in high-resolution dual-comb spectroscopy by making it simple, cost efficient, and compatible with battery operation. Except for the custom-processed graphene-based saturable absorber[33], all components of the system have off-the-shelf commercial availability. A signature feature of our solution is that the generated combs have smooth soliton spectral shapes with almost perfect overlap, hence they do not need any further power-hungry nonlinear broadening mechanisms and spectral filtering to avoid aliasing[21]. Furthermore, the combs' high repetition rates (>140 MHz) are tailored for molecular spectroscopy at low pressure[34] which allows for near-optimal use of optical power per comb tooth. Also, the repetition rate difference is easily tailorable from hundreds of hertz to multi-kilohertz (and even more) by adjusting the intracavity polarization state and by varying the length of a piece of polarization-maintaining fiber. This agility sets it apart from the bidirectional or dual-wavelength configuration, which have weakly tunable repetition rate differences defined by the bidirectional asymmetry of the cavity, and dispersive properties of the fiber, respectively. The accessible spectral range of our system exceeds 1.7 THz within 200 ms of coherent averaging, corresponding to a record-high number of corrected lines for purely-computational phase correction exceeding 11000 here. We prove the high-resolution spectroscopy capabilities of our system on a NIST standard reference material of low-pressure hydrogen cyanide $H^{13}C^{14}N$, which we measure in the $2\nu_3$ overtone band around 1.56 μm in more than 1 THz bandwidth enabled by leveraging a nontrivial computational phase correction algorithm addressing the previous issues of large computational complexity[35,36] with a potential for DCS-platform-independent character[37]. Finally, our single-cavity DCS system itself has been proven to operate in a dual-comb mode over unprecedentedly long timescale of a dozen of hours.

## Results

**Generation of the dual-comb source.** The experimental setup of our SCDC laser is shown in Fig. 1a along with photographies included in Supplementary Figure 1. The all-fiber ring cavity is approximately 1.4 m long, where the dual comb pulses share a common path while propagating in the same direction. Because the cavity is partially birefringent, the nearly-orthogonal pulses will travel at two different group velocities, and hence possess slightly mismatched repetition rates.

The ring cavity includes a 26 cm long piece of Erbium-doped fiber (EDF, Er150) forward-pumped by a 980 nm pump diode. To enable passive mode-locking, we use a graphene/poly(methyl methacrylate) (PMMA)-based saturable absorber on placed onto the facet of an angle-polished fiber connector[33]. At the heart of the cavity is a hybrid component (TIWDM) ensuring unidirectional operation of the laser, which additionally serves as an output coupler for the two combs (10% coupling ratio) and a wavelength division multiplexing (WDM) device for the pump. The birefringence of the cavity is provided predominantly by an 18 cm long section of 980 nm polarization-maintaining fiber (PM980), with a beat-length of ~2 mm, corresponding to an effective refractive index difference $\Delta n_{eff}$ of ~8·10$^{-4}$. To control the repetition rate difference and intracavity polarization state, the PM fiber is preceded by an in-line polarization controller, whereas the rest of the cavity consists of an ordinary single-mode 1550 nm fiber (SMF-28e) (the agility of the repetition rate difference in discussed in Supplementary Note 2). The net cavity dispersion is estimated to be –0.02 ps$^2$, thus the laser operates in the anomalous dispersion regime. Dual-comb lasing requires slight adjustments of the polarization controller, and a sufficient level of optical power pumping the EDF ranging from 110 mW to 121.6 mW provided by the pumping diode with ~350 mW of electrical power consumption in the high-power range. Losses introduced by the TIWDM component from the pump input to output were equal to 8.8%.

**Multi-heterodyne beating.** To enable multi-heterodyne beating on a photodetector and thus observation of dual-comb interferograms, the two combs available at the tap of the output coupler need to be guided through a birefringent optical element. We use a polarizing beam splitter (PBS) preceded by a polarization controller adjusted to maximize the heterodyne efficiency and hence the strength of DCS interferogram, albeit a piece of PM fiber has shown to serve for this purpose as well yet with suboptimal performance. The optical power available on the photodetector was approximately 900 μW, yielding a millivolt-level RF electrical signal low-pass-filtered to 70 MHz and amplified by a low noise amplifier. To record a stream of DCS interferograms we use an 8-bit oscilloscope in high-resolution (oversampling) mode with 11 effective number of bits (ENOB). To simplify the setup, we use a symmetric (collinear) configuration with a single photodetector, where both combs interrogate the sample (AC) which is justified by the smooth shape of the optical soliton spectra for synthetic baseline fitting[38] and negligible frequency difference compared to the molecular line widths probed here (see Supplementary Note 3).

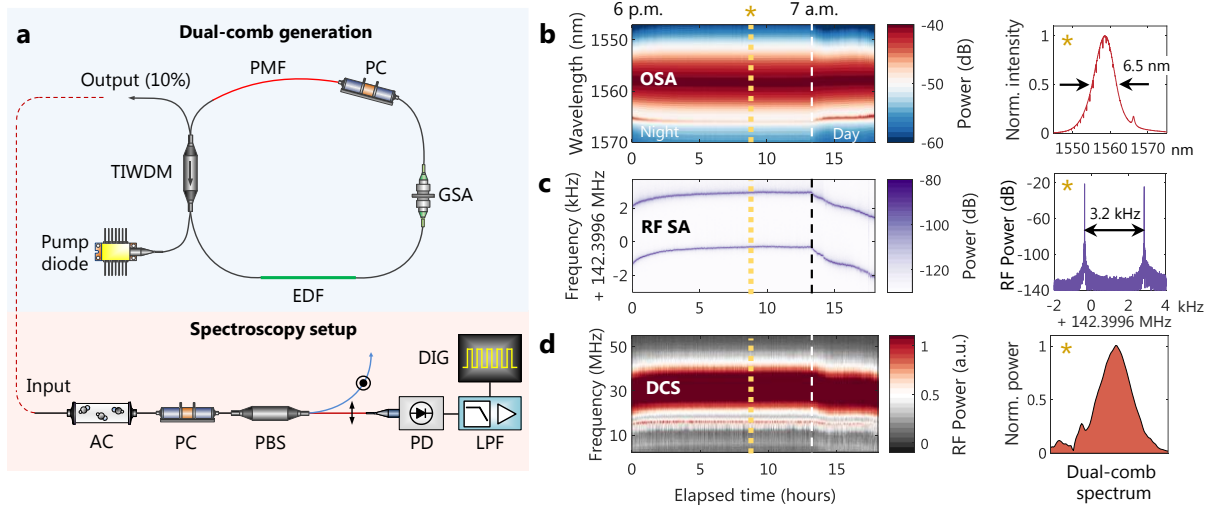

**Fig. 1** Experimental setup for performing dual-comb spectroscopy together with long-term stability of the system. **a** The experimental setup consist of a dual-comb generation block (blue) and spectroscopy setup (orange panel). The dual-comb generation setup includes a polarization insensitive hybrid Tap/Isolator/WDM component (TIWDM), followed by a gain-providing low dispersion erbium-doped fiber (EDF) forward-pumped by a 980 nm pump diode. Passive mode-locking is enabled by a graphene-based saturable absorber (GSA), while the birefringence of the cavity is provided by a 980 nm polarization-maintaining fiber (PMF) preceded by an in-line polarization controller (PC). In the dual-comb spectroscopy setup, the output of the laser feeds a 10-Torr hydrogen cyanide absorption cell. To maximize the heterodyne efficiency on the extended InGaAs photodetector, we use a polarization controller (PC) and a polarizing beam splitter (PBS), however dual-comb interferograms become detectable even after guiding the combs through a piece of PM fiber. **b** Characterization of the optical spectrum over 18 hours of uninterrupted operation. **c** Characterization of the dual-comb RF intermode beat notes. **d** Dual-comb spectrum retrieved from digitized interferograms. The dashed line marked with an asterisk indicates the time instant at which we generated the cross-section plots visible on the right, while that starting at 7 a.m. indicates the beginning of an increased system drift.

**Stability of the dual-comb laser.** One of the main prerequisites for widespread adoption of SCDC lasers outside the laboratory is a demonstration of their ability to sustain in dual-comb lasing mode over long time scales. To date, however, stability tests were performed in intervals ranging from seconds to minutes, which left the question of their long-term operation feasibility unanswered. Initially, we encountered major issues with the reliability of the laser, but those were significantly minimized after we replaced the birefringent part of the ring. We observed that by using a PM fiber optimized for 980 nm (Panda PM 980) with a beat length reduced by 50% compared to PM1550, we can reach a kilohertz-range repetition rate difference with a considerable improvement in long-term stability of the dual-comb source. Fig. 1a, 1b, and 1c plot the optical spectrum, electrical intermode beat note spectrum, and the radio frequency dual-comb spectrum, respectively, measured in an overnight test lasting for 18 hours and displayed in a waterfall form. During that time the laser was left completely free-running, thermally-uncovered and directly exposed to environmental mechanical and thermal fluctuations. For the first 5 hours after dusk, the laser was reaching a steady state related to a temperature change in the laboratory, after which it operated with a negligible drift until morning (7 a.m.). With the morning start of the heating system and routine laboratory activities involving slight mechanical perturbations of the table, a slow drift of the repetition rates and optical spectrum can be observed, albeit the dual-comb operation was sustained until the end of the test interrupted on demand. The shared-cavity configuration ensured that the way the repetition rates evolved throughout the measurement was highly correlated, thus yielding high mutual stability. The side panels of Fig. 1a, 1b show a cross-section of the spectral maps after 9 hours with a smooth 6.5 nm wide soliton optical spectrum mapped to the radio frequency domain with a repetition rate difference of ~3.2 kHz.

**Spectral characteristics.** A major advantage of employing polarization multiplexing in the cavity for dual-comb generation is excellent spectral overlap between the smooth soliton spectra provided intrinsically by the laser[39]. Figure 2 plots the polarization-resolved optical spectra centered around ~1560 nm measured with a resolution of 50 pm using an optical spectrum analyzer together with those retrieved from the radio-frequency interferograms. Due to different losses experienced in the ring cavity, the two polarizations slightly differ in peak intensities (~15%), and in center wavelengths, yet by only ~1 nm, as shown in Fig 2a,b. Both effects, however, have almost no effect on the RF dual-comb spectrum which is the product of the comb teeth intensities rather than the sum. The overall shape of the optical spectrum resembles that of soliton lasers operating in the negative dispersion regime with weak Kelly sidebands located more than a THz from the carrier, whose position was tunable using the intracavity polarization controller. Fig. 2c, illustrates a notable discrepancy in the polarization extinction ratio (PER) of the individual combs. The lower repetition rate RF beat note corresponding to the comb centered around shorter wavelength has a well-defined polarization with a PER of ~30 dB, as opposed to the higher repetition rate comb shifted towards longer wavelengths with a PER of ~10 dB. By adjusting the polarization controller before the PBS, we maximize the strength of the beating signal, which we coherently average and Fourier-transform, as plotted in panels d and e of Fig. 2 to illustrate the aliasing-free mapping of the optical to radio-frequency domains for comparative purposes.

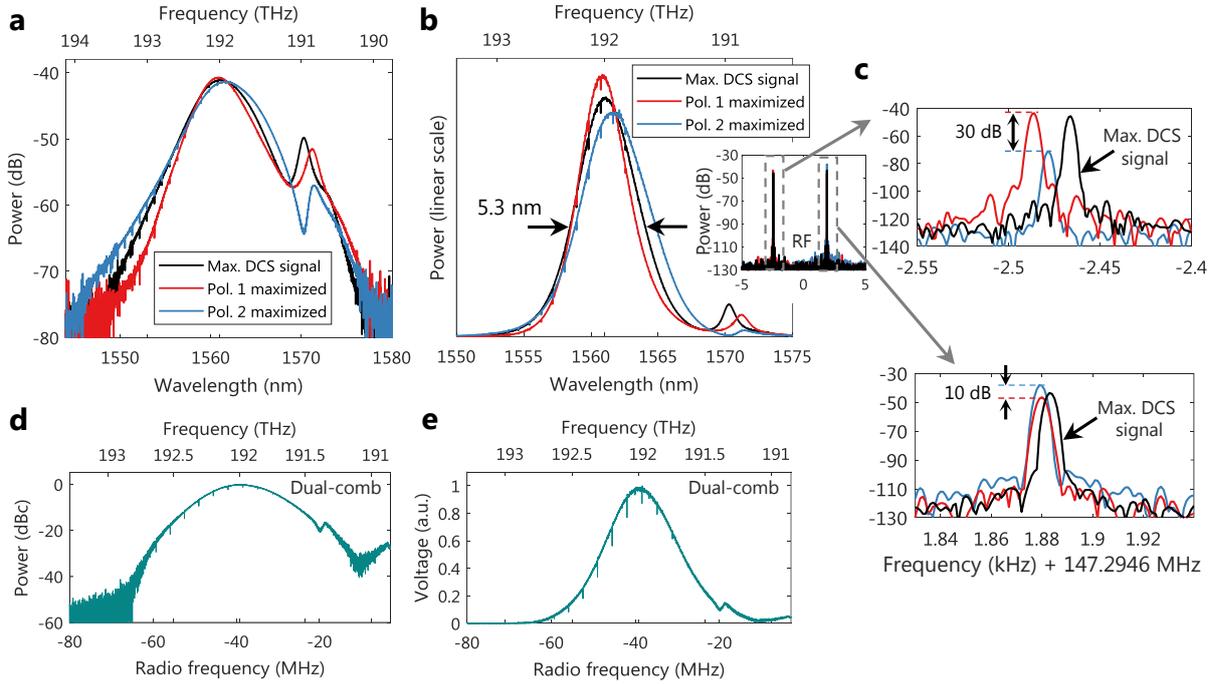

**Fig. 2** Optical and radio frequency spectra of the dual-comb laser after passing through a 10-Torr hydrogen cyanide absorption cell measured at different positions of the polarization controller before the PBS. **a** Optical spectrum in logarithmic scale measured with a resolution of 50 pm. **b** Optical spectrum in linear intensity scale showing the smooth soliton shape and excellent spectral overlap. The detuning of the central wavelengths is ~1 nm. **c** Radio frequency intermode beat notes corresponding to the optical spectra in (a) and (b) at different settings of the polarization controller. The lower-repetition rate beat note corresponds to the shorter-wave-centered comb with a higher polarization extinction ratio (PER) of ~30 dB compared to the longer-wavelength-centered with a PER of ~10 dB. **d** Optical spectrum in a logarithmic scale retrieved from dual-comb interferograms acquired over 200 ms. **e** The same plot as (d) but in a linear intensity scale.

**Linewidth measurements.** For the demanding application of free-running Doppler-limited spectroscopy, it is of utmost importance to carefully characterize the relative and absolute stability of the comb teeth. For this purpose, we optically beat a narrow-linewidth (~200 kHz) 1560 nm continuous wave (CW) laser with two comb teeth, one from each comb[28]. Fig. 3a shows the spectrogram of the low-pass filtered beating signal with visible frequency oscillations of the two RF lines, albeit with a high degree of visual similarity. By mixing the RF lines through a nonlinear operation on the signal (see Methods), we observe that most of these fluctuations are common, and yield a straight-line differential frequency component in Fig. 3b. To fully characterize the efficacy of common phase noise suppression, we calculate the Fourier transform of the time-domain inter-comb mixing signal and measure a relative linewidth between the combs of 200 Hz. This is several orders of magnitude lower compared to the absolute linewidth (coarsely estimated to be 3 MHz) and proves that the laser shows mutual coherence on time scales much greater than $1/\Delta f_r$, as required by computational correction. Such narrow free-running relative line widths have been previously reported only in dual-wavelength configurations[21–25], and may suggest that the large repetition rate difference of several kilohertz is not necessary. Unfortunately, the opposite is true. We estimated, that 80% of the relative beat note power is carried in a 3 kHz bandwidth over 10 milliseconds, and 90% in 6.8 kHz (see Supplementary Note 4). Consequently, to resolve extremely narrow Doppler-limited features, we mismatch the repetition rates by ~4.65 kHz to optimally use the available alias-free RF and optical bandwidths.

To measure the optical (absolute) linewidth of the comb teeth, we analyzed the lower-frequency beat note in Fig. 3a originating from optical beating between the CW laser and a comb tooth close to the RF carrier frequency. Due to its high non-stationarity and rapid frequency fluctuations, we analyzed the beat note in two ways to better visualize the sporadic excursions over a wider frequency range. Figure 3d plots a beat note persistence spectrum, where most of the fluctuations resides in a 6 MHz bandwidth with maximal drifts not exceeding 5 MHz from the center. The conventional frequency spectrum plotted in a linear intensity scale provides an even more optimistic estimate of ~3 MHz. We should underline here, that the retrieved value is of course not as accurate as that measured in an experiment with a hertz-stabilized frequency comb[40]. Instead, it provides an upper bound due to the convolution between the profiles of the sub-MHz-wide CW laser and the comb line. Even though the true comb linewidth is consequently overestimated, it is almost a hundred times narrower compared to the room temperature Doppler linewidths of typical molecular absorbers with low molecular weight, such as $H^{13}C^{14}N$ investigated here.

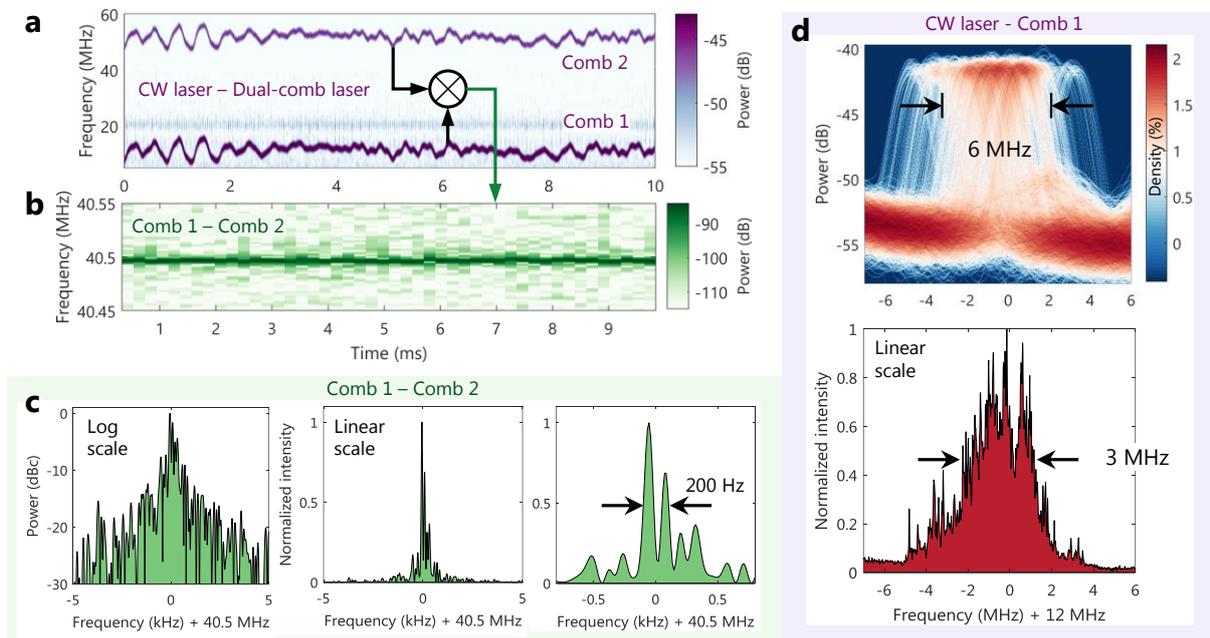

**Fig. 3** Experiment with a narrow-line continuous wave (CW) laser to characterize the absolute stability and mutual coherence of the combs. **a** Spectrogram of the heterodyne signal between the CW laser and two comb teeth, one from each comb. The high visual similarity of the traces indicates a high degree of coherence. **b** Spectrogram of the mixing between the two comb teeth showing a straight line almost free of frequency fluctuations. **c** Fourier transform of the inter-tooth mixing signal plotted in different scales revealing the relative linewidth of 200 Hz within a 100 Hz resolution bandwidth, which is an order of magnitude lower than the repetition rate difference of 4.655 kHz as required for computational correction. **d** Beat note of one comb line and the CW laser plotted in persistence spectrum with logarithmic scale, and the conventional frequency spectrum form in a linear intensity scale. In the persistence spectrum measured with a 1.56 MHz resolution bandwidth, it is clearly visible that for a majority of time the beat note stays in a 6 MHz bandwidth with very sporadic excursions up to <10 MHz. The Fourier-transform-based analysis (which is not well suited for non-stationary signals), provides a line width estimate is on the order of 3 MHz.

**Dual-comb interferograms and phase correction.** Using a fast digital oscilloscope, we acquired a stream of dual-comb-interferograms with an RF repetition rate of $\Delta f_{rep}$=4.655 kHz sampled at 400 MS/s. The temporal structure of a single frame (coherently averaged) is shown in Fig. 4a with a clearly visible extremely short zero path difference (ZPD) burst lasting for approximately 100 ns, which is a time-magnified (compression factor $m$=31642) interval of 3.16 picoseconds related to the cross-correlation between the electric fields of the two combs in the optical domain. It is surrounded by a periodic interferometric modulation induced by the molecular lines of $H^{13}C^{14}N$ known as the free-induction decay, which can be clearly identified here when the temporal signal is averaged. Figure 4b shows that in a single-shot mode, these weak features are buried in noise, hence they necessitate prolonged coherent averaging for noise suppression and reliable spectroscopic assessments. In Fig. 4c one can see that the power spectrum of a single-shot trace shows a large variance with a poor ~10 dB spectral signal-to-noise ratio (SNR), and does not resemble the expected smooth soliton shape. This changes drastically for the coherently average trace, where regularly-spaced narrow lines of hydrogen cyanide become noticeable. The noise floor in the frequency spectrum lowers by approximately 30 dB, being consistent with the number of averaged interferogram frames (see Supplementary Note 5 for evaluation of the coherent averaging efficacy).

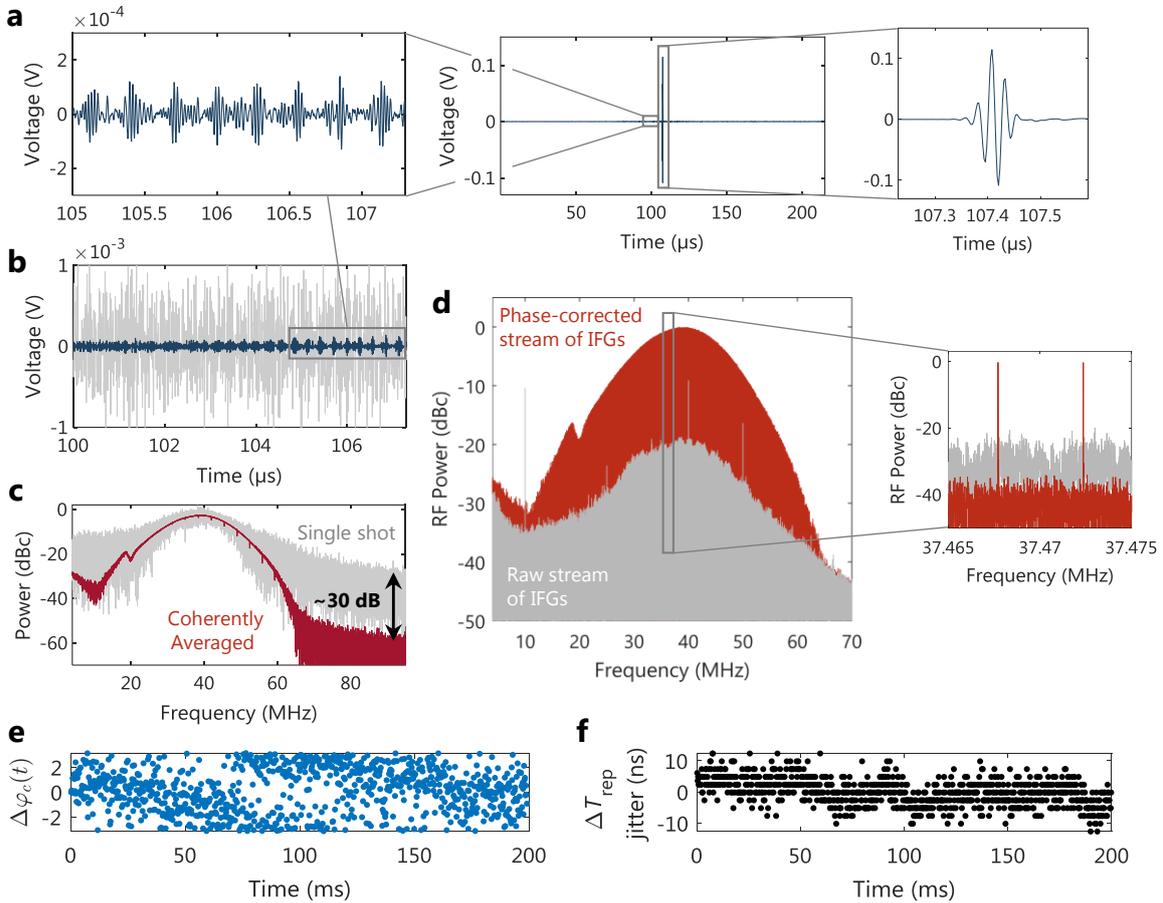

**Fig. 4** Dual-comb interferograms and their Fourier transforms. **a** Time-domain dual-comb interferogram after computationally-enabled coherent averaging. **b** Zoom on interferometric modulation induced by the free-induction decay of HCN molecules plotted atop of a single-shot trace. **c** Power spectra of the single-shot and coherently averaged interferograms. **d** Tooth-resolved RF spectra of the corrected and raw stream of interferograms. **e** Phase increments between the consecutive interferogram frames. **f** Repetition rate difference jitter. 10 nanoseconds of interferogram duration jitter correspond to ~316 femtoseconds of laboratory time.

To perform the time-domain averaging demonstrated in the panels of Fig. 4, we invented a novel computational phase correction algorithm suitable for fast coherent averaging of DCS data. Basically, the RF comb in DCS shows two kinds of instabilities: that of the repetition rate, and that of the frequency offset, which smear out the discrete character of RF lines. The algorithm proposed here corrects for both sources of instabilities in an all-computational way relying only on digitized DCS data from a single detector, which is in stark contrast to the widely adapted CW-laser referencing schemes. Without averaging, the Fourier spectrum of a stream of DC interferograms Fig. 4d turns into a featureless blob with some spurious RF lines originating from harmonics and subharmonics of the digitizer clock. The phase and timing manipulation restores the initially corrupted shape of the spectrum with a Fourier-limited line width and a suppression of the spurious non-comb lines. A detailed description of the procedure is provided in Methods, but for in short we took a radical departure from the cross-ambiguity[36], Fourier-based phase retrieval or Kalman filter[35] paradigms, which hampered the speed of computational correction algorithms. Our time-domain-only solution relies on a constant fraction discriminator (CFD) to track the arrival times of consecutive centerbursts and hence estimate variations in the repetition rate difference (timing jitter), followed by tracking carrier phase and frequency increments between consecutive interferogram frames, results of which are plotted here in Fig. 4e and f. After frame-by-frame resampling (equivalent to adaptive sampling proposed by Ideguchi et al.[40]) followed by phase correction, we obtain a carrier-envelope-offset-free stream of interferograms possible to process in two ways. In the first, they can be coherently co-added to produce a single time-domain trace with an extremely high dynamic range (see Supplementary Note 5). The second approach is to stitch all interferograms just like they were produced by an intrinsically phase-stable dual-comb spectrometer[41], and calculate a tooth-resolved RF spectrum with a linewidth limited by the acquisition time.

In contrast to the well-established amplitude retrieval procedure of calculating zero-padded magnitude spectra combined with peak search, we propose to take a radically different approach with huge computational savings and robustness. It is well-known that the Discrete Fourier Transform (DFT) can provide a non-distorted amplitude spectrum without spectral leakage of finitely sampled data under the assumption that it is sampled at zero crossings. In other words, the frequencies of the signal must coincide with the sampling bins of the DFT. In the coherently averaged trace, this convenience comes for free – the CEO-free harmonic comb has RF lines located exactly at harmonics of $\Delta f_{rep}$, hence we can obtain a tooth-resolved spectrum just by calculating an $N$-point Fast Fourier Transform without windowing, where $N$ is equal to the number of samples in the coherently averaged interferogram. A similar strategy has already been successfully used in ultra-high resolution Fourier Transform Spectroscopy with frequency combs[42], yet its relevance for DCS has not been explored to date. We compared the two amplitude retrieval techniques (see Supplementary Note 6), and found maximum deviations in the per mille range probably attributed to numerical errors in the calculation of the ultra-high resolution FFT with such narrow RF comb lines.

**Allan deviation analysis.** While the phase correction allows to correct for residual phase excursions throughout the measurement, it is always accompanied by the risk that some undersired modulations introduced by the procedure may corrupt the amplitude. We performed an Allan deviation analysis of amplitude on two RF beat notes with different strengths and distances from the center frequency located around 20 MHz, and 40 MHz respectively, as shown in Fig. 5. The dashed lines correspond to uncorrected data, whereas the thick solid lines correspond to data after phase correction. The gain of phase correction becomes noticeable above 10 ms (~47 interferograms), when the amplitude uncertainty of uncorrected data starts to deviate from a decreasing trend. The weak beat note yields an SNR close to 90 after 100 ms of coherent integration, while the strong one reaches a per mille precision regime, which would not have been possible without the digital correction. The nearly 10-fold discrepancy in SNRs between the beat notes is responsible for an increased standard deviation of the spectroscopic fit close to the edges of the dual-comb spectrum.

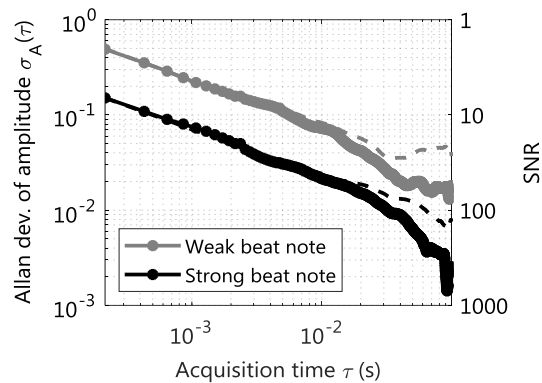

**Fig. 5** Allan deviation of (normalized) beat note amplitude for two different RF lines with (solid line) and without (dashed line) phase correction.

**Doppler-limited spectroscopy of hydrogen cyanide.** Finally, to unequivocally prove the high resolution capabilities of our system and the efficacy of the correction procedure, we measure a molecular sample of hydrogen cyanide ($H^{13}C^{14}N$) at room temperature (22°C) and a low pressure of 10 Torr. In this regime, absorption linewidths are predominantly limited by Doppler broadening. Around the 1560 nm, the Doppler half width at half maximum (HWHM) is assumed to be constant, equal to $\alpha_D$=225(1) MHz, which combined with Lorentzian pressure-broadening coefficients as low as ~10 MHz/Torr renders relatively narrow lines with FWHM Voigt profile linewidths below 600 MHz. In the dual-comb spectrum, it corresponds to approximately 4 lines in a logarithmic absorbance scale, and a bit more in linear transmission.

Because our system is left completely free-running and it is intended to be used without any additional optical components but the oscillator itself, we take advantage of the sharp HCN lines with well-defined frequencies, to anchor the optical frequency axis ($P20$ transition is used here) for spectroscopic assesments. We acquire DCS data in 200 ms (limited by the digitizer), and process it in the aforementioned coherently averaged single-frame mode, where a 85930-point Fourier Transform yields discrete points corresponding exactly to the individual RF comb teeth without any line shape influence. The computational correction of the DCS data on a Core-i7 PC computer with 32 GB of memory takes approximately 24 seconds, yet no significant efforts were put on optimizing the performance of the code written in a high-level Matlab environment. It is, however, real-time compatible thanks to its frame-by-frame architecture, and it should be possible to implement it in a hardware platform such as FPGA.

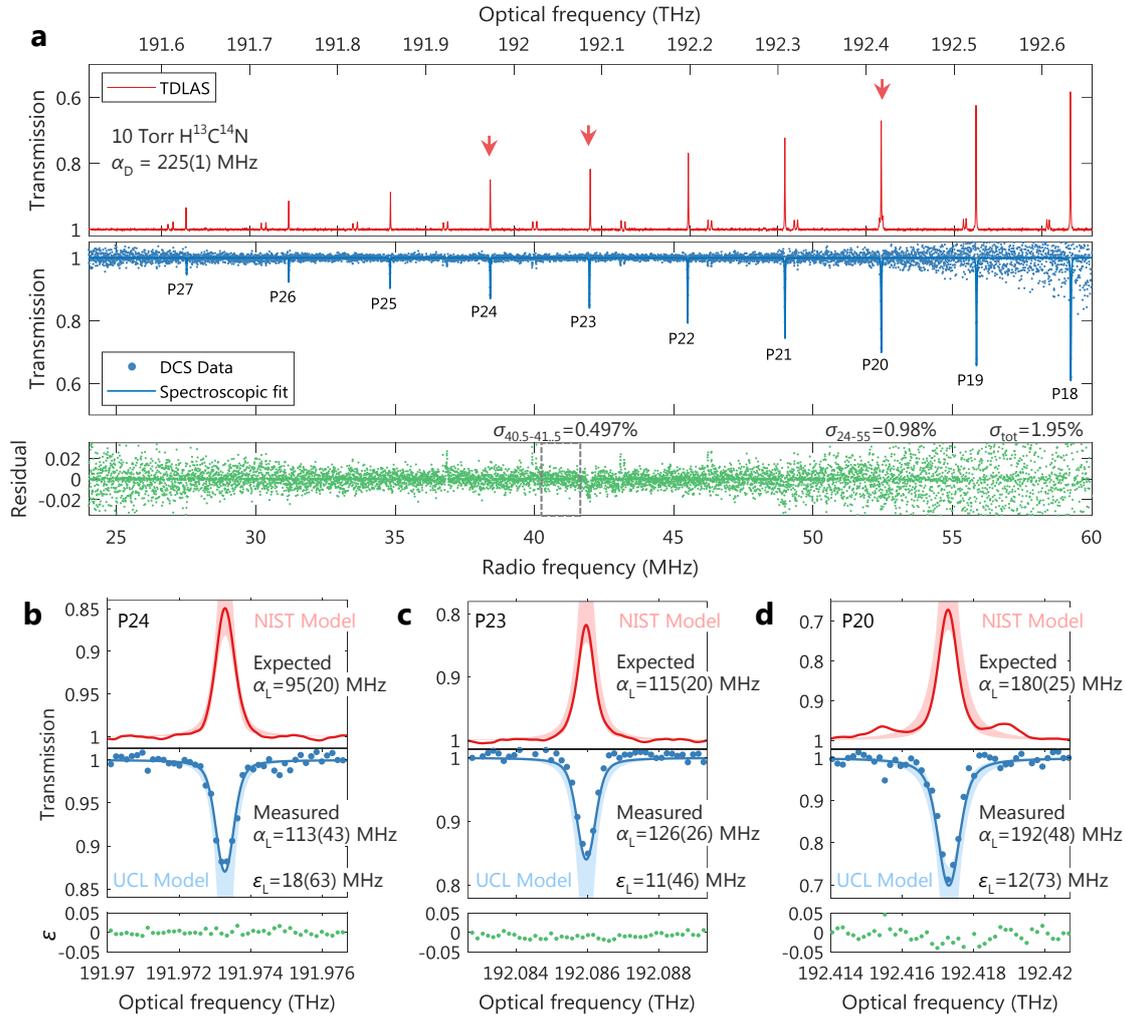

**Fig. 6** Illustration of the quality of the computationally-corrected DCS spectrum of $H^{13}C^{14}N$ at 10 Torr obtained using the dual-comb laser in free-running mode within 200 ms. **a** DCS linear transmission spectrum (bottom) plotted together with a tunable diode laser absorption spectroscopy (TDLAS, top) showing a high degree of mutual similarity. A multi-line Voigt fit to the dominant transitions in the $2\nu_3$ band was obtained from the scatter-point DCS data, while weak interfering hot bands were left unmodeled. The residual calculated from the difference between the DCS data and the fit has a standard deviation of 0.98% in a ~1 THz bandwidth, and 1.95% in total. The peak spectroscopic SNR available in a 30 GHz optical bandwidth (40.5-41.5 MHz in the RF) exceeds 200. **b** P24 line, **c** P23 line and **d** P20 line measured with the dual-comb spectrometer (scattered points) together with a Voigt fit solid line (bottom panel) compared with a tunable laser scan (top panel) and two spectroscopic models. The measured Lorentzian widths are in good agreement with the literature data, and the difference between the expected and measured values lies within the uncertainty of the model. These lines have some degree of weak hot-band interference which may corrupt the fits. Particularly P20 line is surrounded by two unmodeled lines constituting a pronounced oscillatory shape of the residual. Additional weak oscillations on the wings of the absorption lines are caused by Fabry-Pérot etalons from the absorption cell.

To calculate the optical transmission, we leverage the smooth shape of the spectrum and fit a synthetic baseline[38] using a robust parabolic regression method to serve as a zero-gas measurement for spectral deconvolution. We analyze the amplitude rather than power spectra to calculate the transmission because both combs interrogate the sample, as opposed to the widely used dispersive configuration[10]. Figure 6a plots the results of the DCS-based optical transmission measurement together with a multi-line Voigt profile fit based on parameters in Ref.[43] (see Methods for details). Since no comprehensive spectroscopic models are available for this hydrogen cyanide isotope ($H^{13}C^{14}N$), to prove the validity of our results we performed an additional tunable diode laser absorption spectroscopy (TDLAS) experiment using the same cell, wherein the frequency of the high-fidelity tunable CW laser was swept over ~10 seconds, and the intensity was measured using the same photodetector and oscilloscope. For optical axis calibration, this time we use two absorption lines of the *P* branch – *P*24 and *P*20 with vacuum frequencies and pressure shift coefficients derived from the literature[43]. Both transmission spectra are plotted in mirrored panels for comparative purposes and show high degree of mutual similarity.

The quality of the measured DCS data can be accurately evaluated in spectroscopic terms by analyzing the three lines of the *P* branch captured in the spectrum that have been modeled using quantum chemistry methods[44] with confirmed experimental validation[28,43]. In our dataset, we can only model lines *P*24, *P*23, and *P*20 (the only ones available with pressure broadening coefficients), and denote the quantum chemistry-based (ExoMol database[44]) simulation as UCL (University College London), whereas the experimentally-obtained line intensities were used in the NIST (National Institute of Standards and Technology) model[28]. The reason for the simultaneous use of both is that discrepancies in line intensities as large as 11% have been reported[28], and the exact values may still require further experimental verification. The shaded area in Fig. 6b, c, and c corresponds to $2\sigma$ (95%) confidence intervals associated with combined uncertainties of the broadening coefficients, line intensities, and gas pressure (see Methods). Also, based on the Voight fits we compare the fitted Lorentzian half-widths $\alpha_L$ for the three available lines to pressure broadening coefficients reported in Ref.[43]

Overall, we obtain excellent agreement between the Voigt fits and DCS data (also including the lines without a spectroscopic model). Peak intensities of the lines follow the same increasing trend as in the TDLAS measurement and reach values different from the TDLAS measurement by ~2% of transmission scale in the worst cases, possibly attributable to slight nonlinearities of the photodetector causing a difference in response at different optical power levels. The modeled lines *P*24 and *P*23 possess generally lower oscillations on the wings compared to *P*20 due to their much weaker hot-band interference. The presence of those weak unmodeled transitions adjacent to strong lines necessitates to limit the range of data around the peaks for the fitting routine[43] (see Methods), and contributes to sharp dips in the residual due to their exclusion from the spectroscopic model. The fitted Voigt profiles are consistent with the theoretical UCL and NIST models and lie within their $2\sigma$ confidence intervals. The same is true with respect to the fitted Lorentzian linewidths, which match the literature values within uncertainties associated with them.

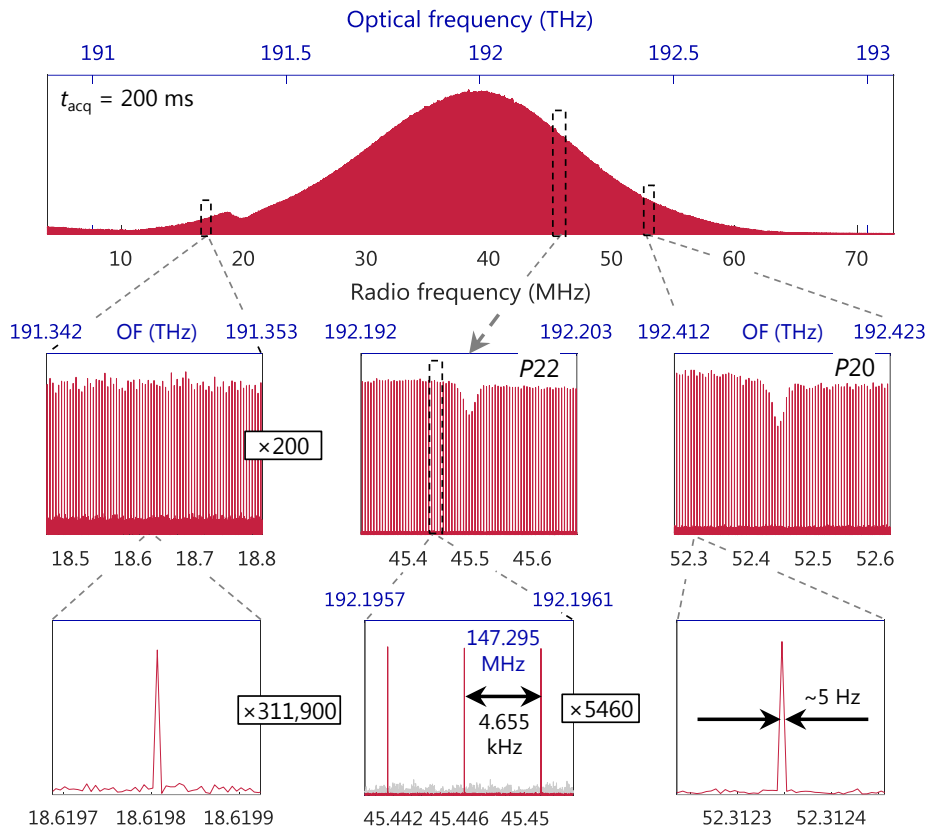

**Fig. 7** Computationally corrected dual-comb amplitude spectrum in linear scale covering 1.7 THz in the optical domain with resolved 11000 RF comb lines plotted with different magnification factors on the frequency axis. The central panels include two transitions of the *P* branch: *P*22 and *P*20 with different levels of signal-to-noise ratio. The bottom panel with three comb teeth includes additionally the spectrum without phase correction (gray blurred trace) with spectroscopic information dispersed over a large RF bandwidth.

This proves that such a simple laser combined with the computational algorithm renders quality spectra in the Doppler-limited resolution range. Within 200 ms, we achieve better than 1% precision in a ~1 THz optical bandwidth, and less than 2% over the full span of ~1.14 THz with, based on the fit residual. It corresponds to a noise equivalent absorption (NEA) of $4.4 \cdot 10^{-3}$ Hz$^{-1/2}$ and $8.7 \cdot 10^{-3}$ Hz$^{-1/2}$, respectively, whereas the figure-of-merit (FOM) defined as SNR at 1 second of averaging (NEA$^{-1}$) multiplied by the number of resolved lines $M$ in a given bandwidth[45] yields $1.5 \cdot 10^6$ Hz$^{1/2}$ in 1 THz when $M$=6660, and $8.9 \cdot 10^5$ Hz$^{1/2}$ in the full 1.14 THz window when $M$=7734. The maximum spectroscopic SNR obtainable with our system within 0.2 s exceeds 200 in a 30 GHz optical span around 192.05 THz.

As a final demonstration of the computationally-enabled spectrometer performance, we plot an RF tooth-resolved dual-comb spectrum covering ~1.7 THz mapped on more than 11000 lines. This is the largest number of self-corrected lines of a free-running dual-comb system reported to date (i.e. without using any CW lasers, $f$-$2f$ interferometry, and additional optoelectronic components[40]). Fig. 7 plots the spectrum with different magnification factors of the frequency axis, and confirms that even components located thousands of lines away from the carrier have a comb structure without any increase in line width or appearance of a pronounced noise pedestal.

## Discussion

Our demonstration of free-running high-resolution spectroscopy using an ultra-simple, all-fiber single-cavity dual-comb laser (AFSCDCL) relying on polarization multiplexing inside a birefringent cavity bodes for a wide range of possible applications in the laboratory and industrial environment. It paves the way for low-drive-power and cost-effective dual-comb spectrometers with an easily-tailorable repetition rate difference, which accompanied by a high repetition rate exceeding 140 MHz, permit a further extension of optical bandwidth without optical aliasing in DCS. We prove that operation of a AFSCDCL on a dozen-hour time scale is possible, albeit without any indication that the ultimate limit is reached. By merging the unprecedentedly simple cavity design utilizing just a few basic fiber components with our novel all-computational phase correction algorithm, we can effectively suppress residual phase fluctuations of dual-comb interferograms, thus rendering high-quality molecular dual-comb spectra with tooth-resolved lines, sub-percentage precision in a THz bandwidth measured in a fraction of a second, and corrected within a user-acceptable time of less than half a minute. The processing speed, however, is not a limitation *per se* and can be easily improved by implementing the algorithm in a hardware platform or by lowering the computational payload through an external quadrature demodulator. The excellent agreement between the measured spectroscopic data in the Doppler-limited regime and spectroscopic models together with a narrow-linewidth tunable diode laser scan prove that computationally-corrected free-running spectrometers are suitable for high-resolution spectroscopy applications, and that more than ten thousand teeth in the dual-comb spectrum can be corrected without noticeable degradation of lines away from the carrier. Finally, by exploiting the algorithmic removal of the carrier-envelope-offset frequency from DCS interferograms, we propose to calculate a sparsely-sampled Fourier transform directly on the averaged interferogram. It results in considerable computational savings compared to the conventionally obtained apodized high-resolution spectrum after multi-fold zero-padding followed by peak searching. The gain of computational coherent averaging reached here 99.3% of the theoretical, whereas amplitude deviations between the sparsely-sampled and high-resolution Fourier spectra reached at most a per-mile regime.

Although our free-running dual-comb source will never reach the precision and accuracy of the highly-stabilized and absolutely-referenced frequency combs, optical frequency fluctuations occurring on typical measurement time scales do not preclude measurements of high-quality low-pressure molecular spectra. With future developments and active environmental stabilization of the cavity, we anticipate further improvements of optical bandwidth and system stability. Most importantly, to exploit the full spectroscopic potential of the developed AFSCDCL enhanced by the flexible coherent averaging procedure, we will focus on shifting the comb wavelengths through nonlinear conversion to other spectral regions such as the mid- or far-infrared .

**Data availability.** The data that support the plots within this paper and other findings of this study are available from the corresponding author upon reasonable request.

**Code availability.** The computational coherent averaging Matlab program used to compute the tooth-resolved RF spectrum is available from the corresponding author upon reasonable request.

## Methods

**Detailed dual-comb generation setup.** In the dual-comb generation setup, we used a fiber-coupled pump diode with a center wavelength of 976 nm supplied with 250 mA of current and a voltage bias of 1.43 V (SK17511049) provided by a laser driver (CLD1015, Thorlabs). Under such bias conditions, the diode emitted ~119 mW of optical power corresponding to ~33% of wall-plug efficiency, and this parameter predominantly leaves room for improvement of the power budget in future implementations. The pump light was coupled to the ring oscillator through a single-stage Tap/WDM/Isolator device (TIWDM-980/1550S-10%-250, Opneti) introducing 8.8% of losses from the pump input port to the output. The TIWDM served also as a 10% output coupler. It was spliced to a 26 cm long piece of low-dispersion erbium doped fiber (EDF150LD, OFS Optics) with a peak absorption of 150 dB/m and a dispersion of −16 ps/(nm·km), further followed by a graphene-based saturable absorber (SA). To fabricate the absorber, 20 layers of graphene were deposited on a poly(methyl methacrylate) (PMMA) substrate using procedures described in Ref.[33], and next the substrate with the graphene was transferred onto the facet of an angle-polished fiber connector. Finally, the absorber was sandwiched between a complementary connector so that its center was located ~72 cm from the output tap (measured to the center of the hybrid component). The cavity included also an in-line polarization controller (F-POL-IL, Newport) preceding an 18 cm long piece of panda-type polarization maintaining fiber optimized for 980 nm (PM980, Nufern) located 28 cm away from the hybrid component. The total cavity length was ~1.40 m, and except for the PM section, it utilized an ordinary 1550 nm single mode fiber (SMF-28e, Corning). The use of the 980 nm PM section rather than optimized for 1550 nm was dictated by a nearly 50% decreased beat length, which enabled stable dual-comb operation despite the short cavity length allowing to reach higher repetition rates more suited for molecular spectroscopy. The cross-polarized dual-comb light was collected directly from the tap of the hybrid device without any additional optical or optomechanical components. The mode-locked operation of the laser was only possible when the graphene/PMMA composite was placed onto the fiber connector. In the absence of the composite, we did not observe any sign of mode synchronization through the weak artificial saturable absorber realized by the piece of PM fiber and polarization controller. The polarization multiplexing mechanism is in fact strongly supported by the polarization insensitive graphene-based saturable absorber.

**Detailed dual-comb spectroscopy setup.** The output of the dual-comb oscillator was guided to a cascade of two fiber-coupled, 16.5 cm long absorption cells containing hydrogen cyanide ($H^{13}C^{14}N$) at a low pressure of 10 Torr. Next, it was followed by an in-line polarization controller (Fibrepro, PC1100) and a fiber-based polarizing beam splitter (PBS, 1 SM, 2 PM ports, >20 dB extinction ratio). One of the fiber outputs was connected to a fast extended InGaAs photodetector (DCS-50S-39, Discovery Semiconductors) to produce a radio-frequency (RF) dual-comb interferogram. Its electrical output was split into two paths with a resistive RF power splitter for simultaneous measurement of the dual-comb RF intermode beat notes with an RF spectrum analyzer (EXAN9010A, Agilent) and acquisition of dual-comb interferograms, yet in an actual device that splitting may not necessary and served only for diagnostic purposes, just like a measurement of the optical spectrum using an optical spectrum analyzer (AQ6375, Yokogawa, 50 pm resolution) from the other port of the PBS. It should be underlined, however, that in the polarization-resolved characterization (Fig. 2 of the main manuscript), the optical spectrum RF spectra were measured from the same PBS output after splitting with a 50/50 PM fiber coupler.

The electrical output of the photodetector that was used for measuring dual-comb interferograms required electrical amplification and filtering to avoid RF spectral aliasing. We used a home-made 70 MHz LC-type low-pass filter followed by a 40 dB low-noise (0.8 dB noise figure) RF amplifier (PE15A1012, Pasternack), and another low-pass filter prior to sampling the electrical signal with a fast 8-bit oscilloscope (Infiniium DSO90604A, Agilent). We used the oscilloscope in oversampling (high vertical resolution) mode by taking advantage of its high sampling rate (20 GS/s) to acquire digital interferograms at 400 MS/s with 11 effective number of bits (ENOB).

**CW laser experiment.** To characterize the relative and absolute linewidths of the dual-comb laser, we optically heterodyned a selected portion of the dual-comb spectrum with a narrow-linewidth continuous wave (CW) laser (Ando AQ4320D, ~200 kHz linewidth). The fiber-coupled source operating at a center wavelength of ~1560 nm and an optical power of 1 mW was combined with comb lines lying in the central part of the optical spectrum filtered using a home-made tunable optical band-pass filter. We used a non-PM 90/10 optical coupler to guide most of the low-intensity dual-comb light to the 90% coupling ratio port, whereas the 10% port was used for the CW laser acting as a strong local oscillator. The optical beating signal was measured using the previously mentioned extended InGaAs photodetector connected to the oscilloscope (together with the instantaneous repetition rate signal and dual-comb interferograms from a separate detector unit). The CW laser mixing signal was filtered and amplified using the same RF circuitry as in the Detailed Spectroscopy Setup. After acquisition, the measured signal was subject to line width analysis.

First, we generated a waterfall plot (RF spectrogram in Fig. 3a) of the beating signal between two comb lines, one from each comb, and the CW laser, which was calculated with a temporal resolution of 20 μs and an intensity threshold of -55 dB. It reveals highly correlated fluctuations of the two comb lines. From a dual-comb spectroscopy perspective, of particular interest is the relative comb linewidth, hence to characterize it, we digitally squared the CW laser mixing signal to induce self-mixing between the two isolated comb lines, thus giving rise to a strong RF line located exactly at the difference frequency between them. The line was so narrow that for the generation of the difference-frequency spectrogram in Fig. 3b, we picked a frequency resolution of 3 kHz, and 70% overlap between the periodogram blocks at the expense of a relatively sparse temporal grid. To ultimately prove the high coherence between the two RF lines, we calculated the Fourier Transform on the full-length self-mixing signal, which yielded the 200 Hz relative line width (at 100 Hz of resolution bandwidth) quoted in text.

The estimate of the absolute line width of the comb lines was obtained in two ways. First, we calculated the persistence spectrum of the lower-frequency beat note centered at ~12 MHz in a bandwidth of ~14 MHz with a frequency resolution of 1.56 MHz, and a temporal resolution of 20 μs, as plotted in the top panel Fig. 3d. This type of plot is well-suited for analysis of non-stationary signals, i.e. such that change frequency characteristics over time. Sporadic excursions over a wider frequency range, as well as the bandwidth of long-time persistence are clearly visible in the figure. For fair comparison, however, we perform a conventional Fourier-based analysis on the entire acquired signal, as plotted in the bottom panel of Fig. 3d. It should be noted, however, that the estimated linewidth is in fact a convolution of the profiles of the comb lines and the CW laser, hence the quoted number is inevitably overestimated. Nonetheless, such megahertz-scale fluctuations can be neglected when compared to Doppler linewidths of typical molecular transitions at room temperature.

**All-computational phase correction algorithm.** Dual-comb interferograms are predominantly corrupted by two kinds of noise: additive amplitude noise and multiplicative phase noise. Provided that one can deal with the latter, amplitude fluctuations responsible for uncertainty of the spectral transmittance can be minimized through prolonged averaging. Unfortunately, phase noise cannot be suppressed using trivial techniques, requiring sophisticated carrier and timing recovery algorithms instead. While such are widely developed for wireless network modulation schemes, the field of computational multiheterodyne spectroscopy[35] is just emerging. And it is not the correction that is difficult *per se*, but the retrieval of phase and frequency fluctuations. Unlike techniques relying on external CW lasers to extract the correction signals, our solution is all-computational and operates on raw time-domain interferograms. Rather than relying on extremely effective yet computationally demanding cross-correlation or cross-ambiguity functions[36] or the Extended Kalman Filter (EKF)[35] to simultaneously extract the RF carrier frequency $f_c$ and RF pulse repetition rate $\Delta f_{rep}$, we took a radically different approach. We decomposed the problem into two sub-problems: carrier-frequency-independent estimation of the interferogram's time-of-arrival followed by frame-to-frame carrier phase tracking.

We take advantage of the characteristic temporal structure of the dual-comb signal. Because there is a strong centerburst in the interferogram, we can use it as a trigger to measure the quasi-instantaneous RF repetition rate $f_{rep}$. Unfortunately, ordinary threshold (leading edge) triggering introduces a considerable timing jitter and shows high sensitivity to carrier phase. To make the triggering scheme virtually insensitive to carrier parameters, we can calculate the envelope of the signal either by quadrature demodulation of the RF dual-comb signal or by applying the Hilbert Transform to obtain an analytic representation. In a real-time platform, a finite-impulse-response (FIR) approximated Hilbert transform filter can be implemented, which is now supported by most hardware architectures. In principle, this operation is similar to amplitude demodulation of a signal. Mathematically, a real (passband) dual-comb signal $y(t)$ with an instantaneous carrier frequency $f_c(t)$ can be converted into the analytical representation through

$$y_a(t) = y(t) + i \cdot \mathcal{H}(y)(t) = A(t) \cdot \exp\left(i\left(\varphi_c(0) + 2\pi \int_0^t f_c(\tau)\mathrm{d}\tau\right)\right), \tag{1}$$

where $A(t)$ is the instantaneous amplitude of the envelope, $\mathcal{H}$ is the Hilbert transform, and $\varphi_c(0)$ is the initial phase. By taking the absolute value of $y_a(t)$, we obtain a carrier-frequency-independent signal, which after low-order low-pass (FIR) filtering yields a good estimate of the true instantaneous amplitude. Alternatively, one can use hardware-based quadrature (IQ) demodulation to obtain the same. That signal is next fed to a constant fraction discriminator (CFD) block, which in contrast to leading-edge filtering does not introduce timing jitter and is more robust than ordinary maximum peak trigger failing to be accurate for signals without a sharp maximum. This technique is widely used in scintillation counters and relies on triggering at a level being a fraction of the peak height with a theoretical zero time walk. The CFD basically consists of an ordinary delay, negation, summation and zero crossing detector block (see Supplementary Note 7). The input signal is divided into two, one of which is inverted, time delayed by $\tau_D$, and added to the other. The resulting waveform has a well-defined zero crossing suitable for triggering. In other words, we find such $t$ around the centerburst where the sign of $A(t) - A(t - \tau_D)$ changes from positive to negative. Of course, the *a priori* known delays introduced by FIR filtering and the delay block of the CFD should be compensated for accurate determination of the time of arrival. Next, we divide the signal into individual interferogram frames, and derive the discrete (frame-by-frame) repetition rate $\Delta f_{rep}[i]=1/\Delta T_{rep}[i]$, followed by adaptive resampling[40,37] using a cubic polynomial to ensure the equal duration of the frames. In the repetition correction we assume that the repetition rate changes linearly from the center of the interferogram on the left $\Delta f_{rep}[i-1]$ to the center of the corrected frame with a repetition rate $\Delta f_{rep}[i]$, and next changes linearly to the center of the interferogram on the right with $\Delta f_{rep}[i+1]$, which yields a continuously varying $\Delta f_{rep}(t)$. Next, we provide an additional phase shift[46] induced by the timing jitter $T_{jitter}(t)$, which is obtained by multiplying the complex resampled interferogram frame $y_r(t')$ by $\exp(i2\pi f_c T_{jitter}(t'))$, thus yielding $y_{rc}(t')$. Now all the comb teeth share common frequency fluctuations, as visible in the spectrum of the digital difference frequency (DDFG) signal[37] (see Supplementary Note 7, Supp. Fig. 8), and thus any dependence of phase noise characteristics on the mode order number is virtually eliminated. The only remaining fluctuation is common phase noise. Rather than calculating the absolute phase around the center of each interferogram, we calculate the phase increment between the corresponding samples around the center of the current ($i$) and previous ($i-1$) interferogram frame. For the discrete time case that is $y_{\text{int}}[i,n] = y_{\text{rc}}[i,n]y_{\text{rc}}^*[i-1,n]$, where the asterisk denotes complex conjugate, and $n$ is the sample number. Next, we obtain the mean of the complex samples $y_{\text{int}}[i,n]$, and retrieve the (mean) discrete-time advancement of phase between two consecutive interferogram frames $\Delta\varphi_c[i] = \arg\{\bar{y}_{\text{int}}[i]\}$. Finally, a linear phase correction vector $\varphi_{\text{cv}}$ is constructed by the same way as in the repetition rate case (phases in the center of the neighboring left $\Delta\varphi_c[i-1]$, current $\Delta\varphi_c[i]$, and neighboring right $\Delta\varphi_c[i+1]$ interferograms are known, and they are assumed to vary in a linear fashion). Phase correction of the of $i$-th interferogram frame therefore given by

$$y_{\text{cor}}[i,n] = y_{\text{rc}}[i,n] \cdot \exp\left(-\mathrm{i}\left(\varphi_{\text{cv}}[i,n] + \left(\sum_{k=1}^{i-1}\Delta\varphi_c[k]\right)\bmod 2\pi\right)\right). \tag{2}$$

Note that this phase correction ensures equal phases of all interferogram frames, hence it turns the RF comb into a harmonic one, i.e. without a carrier-envelope offset. Despite the apparent complexity, the algorithm operates in an all frame-by-frame architecture with no significant computational requirements. Repetition rate correction is basically a linear (or cubic) interpolation problem, whereas phase carrier extraction and correction are nothing but a calculation of the phase angle and complex multiplication supported by the CORDIC algorithm. Consequently, the extraction of the correction parameters is fast and bodes for future real-time implementations.

The proposed digital phase correction method, however, has several limitations, of which the most critical is the necessity of providing sufficient mutual coherence between the combs on a $1/\Delta f_{\text{rep}}$ scale. In other words, large and regular carrier phase jumps beyond $\pi$ may lead to incorrect spectroscopic results. Also, an accumulation of numerical errors may lead to a drifting carrier frequency and phase. It is possible to address this potential issue by introducing a second phase correction step comparing the current frame with the first acquired interferogram or the average of those already accumulated. Another challenge is the discrete nature of acquired samples implying roundoff errors in the zero-crossing detection for repetition rate estimation because the result must be an integer number. Higher sampling rates or digital interpolation filters will definitely improve this procedure. Finally, the proposed method improves only the linewidth of RF comb lines, while leaving optical domain fluctuations intact.

**Spectroscopy of HCN.** The measured isotope of hydrogen cyanide $H^{13}C^{14}N$ has proven its usefulness for precise wavelength calibration of optical instruments in the near-infrared, yet it is still lacking a comprehensive spectroscopic line-by-line model, as opposed to the most-abundant isotope $H^{12}C^{14}N$ conveniently available in the HITRAN[47] database. To demonstrate the precision and accuracy of our spectrometer, we compared the dual-comb measurement fit with a tunable diode laser spectroscopic scan (TDLAS), and the two spectroscopic models of the available *P*-branch transitions as described in the text. We simulated the transmission spectra using formulae in Ref.[47] for the three available lines modeled using a Voigt profile at a temperature of 296 K with theoretical line intensities from the ExoMol database[44], and those obtained experimentally[28], whereas the pressure broadening coefficients common for both models were obtained from Ref.[43]. To account for the isotopic ratio of HCN in the cell, previously reported line intensities were scaled by the natural isotopic abundance from HITRAN[47]. In the simulation, however, we had to account for the total uncertainty budget from numerous contributors of the model, which yielded shaded-area curves representing $2\sigma$ (95%) confidence intervals. Here, the quoted uncertainty is the extended uncertainty with a coverage factor of $k$=2 (uncertainty is $\pm 2\sigma$) induced by the dominant sources of uncertainty: pressure of gas in the cell, pressure broadening coefficients, and line intensities, whereas the molar fraction, isotopic purity, cell length, and temperature were assumed to be known exactly for simplicity.

In the Voigt profile fitting routine, we used the nonlinear trust region reflective algorithm and limited the number of points being fit to the central part of 20 points around the peak: 10 on each side, which corresponds to approximately 7 Doppler widths. This was dictated by the presence of weak adjacent lines from hot-band transitions, which would otherwise corrupt the fit[43]. The Doppler widths were known a priori and fixed, while the line intensity, Lorentzian width, and center position were free parameters for the fitting routine for each line. To obtain the uncertainties of the Lorentzian line widths from the fits, we calculated them from the Jacobian estimate and the variance of residuals.

**Acknowledgements**

We warmly thank Dr. Adam Fleisher of NIST for useful comments regarding the uncertainty budget of spectroscopy of HCN, and Rohde & Schwarz Poland for lending us the real-time radio-frequency analyzer. This work was partially supported by National Science Centre (NCN, Poland) under the grant no. UMO-2014/13/D/ST7/02143, the Statutory Funds of the Chair of EM Field Theory, Electronic Circuits and Optoelectronics (Mloda Kadra 0402/0157/18) and by the Research Foundation Flanders (FWO) under Grant no. EOS 30467715. Ł.A. Sterczewski acknowledges support from the Foundation for Polish Science within the START Program.

**Author contributions**

Ł.A.S. and J.S. conceived and performed the experiments. Ł.A.S. developed the computational correction algorithm, analyzed the data, and performed the spectroscopic simulations. Ł.A.S prepared the manuscript with input from J.S. and in discussion with all authors. A.P. and W.K. fabricated the saturable absorber. J.S. initiated and supervised the project.

**Additional information**

**Competing interests:** The authors declare no competing interests.